\documentclass[nofootinbib,superscriptaddress,reprint,onecolumn,notitlepage]{revtex4-1}
\usepackage{amssymb}
\usepackage{amsmath}
\usepackage{graphicx}
\usepackage{array}
\begin{document}

\title{Rastall's Cosmology and its Observational Constraints}
\author{J\'ulio C. Fabris, Oliver F. Piattella, Davi C. Rodrigues}
\affiliation{Departamento de F\'{\i}sica - Universidade Federal do Esp\'{i}rito Santo - Brazil}
\author{Mahamadou H. Daouda}
\affiliation{D\'epartement de Physique - Universit\'e Abdou Moumouni de Niamey - Niger}

\begin{abstract}

The Rastall's theory is a modification of General Relativity touching one of the cornestone of gravity theory: the conservation laws.
In Rastall's theory, the energy-momentum tensor is not conserved anymore, depending now on the gradient of the Ricci curvature. In this
sense, this theory can be seen as a classical implementation of quantum effects in a curved background space-time. We exploit this structure in
order to reproduce some results of an effective theory of quantum loop cosmology. Later, we propose a model for the dark sector of the universe.
In this case, the corresponding $\Lambda$CDM model appears as the only model consistent with observational data.

\end{abstract}

\maketitle

\section{Rastall's theory}

The Rastall's proposal is based on the following observation: The conservation laws are tested effectively only on flat spacetime \cite{rastall}. Hence, in a non-flat spacetime, there is room for generalizations of the usual
conservation laws. The choice made by Rastall to introduce a modification of the usual conservation law is:
\begin{eqnarray}
\label{cons}
{T^{\mu\nu}}_{;\mu} = 0 \quad \Rightarrow \quad {T^{\mu\nu}}_{;\mu} = \frac{(\lambda -1)}{16\pi G}R^{;\nu}.
\end{eqnarray}
with $\lambda$ a constant. When $\lambda = 1$, General Relativity, with its usual conservation law, is recovered.
\par
Today, a new motivation to Rastall's theory can be given. Since the violation of the usual conservation law is related to the gradient
of the curvature scalar, such violation can be associated to quantum effects in curved spacetime. In fact, the expression (\ref{cons}), in a particular form, appears in the
so called gravitational anomaly, a anomaly in the usual gravitational equations due to quantum effects \cite{anomaly}. According to this remark, the Rastall's theory can be considered
as a classical, effective implementation of quantum effects due to fundamental fields living in a curved spacetime.
\par
The field equations of the Rastall's theory can be written alternatively as follows:
\begin{eqnarray}
R_{\mu\nu} - \frac{1}{2}g_{\mu\nu}R &=& 8\pi G\biggr\{T_{\mu\nu} - \frac{\gamma - 1}{2}g_{\mu\nu}T\biggl\},\\
{T^{\mu\nu}}_{;\mu} &=& \frac{\gamma - 1}{2}T^{;\nu}.
\end{eqnarray}
In these equations $\gamma$ is a dimensionless parameter related to the previous one, $\lambda$, by the relation,
\begin{equation}
\gamma = \frac{2 - 3\lambda}{1 - 2\lambda}.
\end{equation}
When $\gamma = 1$, General Relativity, with the usual conservation law, is recovered.
\par
Is this theory well justified theoretically? First of all, there is in principle no Lagrangian formulation leading to the Rastall's field equation. But this is strictly true only at the context of the Riemannian geometry. Possible exterions (e.g., Weyl geometry) may
cure this possible drawback. Moreover an introduction of an "external field" may lead to a Lagrangian formulation even in the context of the Riemannian
geometry.
\par
In fact, consider the following Lagrangian:
\begin{eqnarray}
L = \sqrt{-g}\biggr\{R + \phi(R - \Lambda)\biggl\} + L_m,
\end{eqnarray}
where $L_m$ is the usual matter Lagrangian, $\Lambda$ is an external constant field, and $\phi$ is a Lagrange multiplier.
Such modification of the usual Hilbert Lagrangian of General Relativity in order
to obtain the Rastall's field equation.
The parameter $\gamma$ is connected with $\Lambda$.
\par
Moreover, besides the initial motivations for the Rastall's theory, there are others possible connections with
effective quantum models.
As discussed in references \cite{anomaly,birrell}, quantum effects in curved spacetime lead to violations of the usual conservation law of the energy-momentum
tensor.
This phenomena is usually known as {\it gravitational anomalies}. Such gravitational anomaly may be very important, for example, in the computation of the Hawking radiation of a black hole \cite{glauber}.
\par
For example, consider the effective two-dimensional metric of a black hole:
\begin{eqnarray}
\label{met1}
ds^2 = f(r)dt^2-f(r)^{-1}dr^2\quad.
\end{eqnarray}
The absence of ingoing models leads to \cite{wil},
\begin{eqnarray}
\label{eang}\nabla_\mu T^\mu_{\nu}(r_H)\equiv\Xi_\nu(r)\equiv\frac{1}{\sqrt{-g}}\,\partial_\mu N^\mu_\nu\quad,
\end{eqnarray}
where,
\begin{eqnarray}
\label{ano}
N^\mu_\nu=\frac{1}{96\pi}\,\epsilon^{\beta\mu}\partial_\alpha\Gamma^\alpha_{\nu\beta}\rightarrow N^r_t=\frac{1}{192\pi}\bigg(ff''+f'\,^2\bigg)\quad,
\end{eqnarray}
with $\epsilon^{01}=\epsilon^{tr}=1$ and ($'$) is a derivative with respect to $r$.
\par
Such interpretation of the Rastall's theory as an classical, effective implementation of quantum effects leads to many interesting applications, as we will see in what follows. First, we will use the Rastall's theory to implement a classical version of the Loop quantum cosmology effective equation, showing that singularity-free solutions emerge in that classical context.
Second, we will apply Rastall's theory to the description of the present universe. It will be shown that the corresponding $\Lambda$CDM model is unique when a confrontation with
observations is made.

\section{Application I: LQC}

 Among the many approaches to construct a Quantum Theory of Gravity, there is the
{\it Loop quantum gravity}, based on the use of the connections as the fundamental quantities \cite{lqg}.
A quantum cosmological model can be constructed from LQG, called {\it Loop quantum cosmology}, LQC.
One of the predictions of LQC is the existence of an upper critical density $\rho_c$ leading to the avoidance of the cosmological singularity. This prediction is encoded in a modification of the Friedmann's equation such that \cite{lqc},
\begin{eqnarray}
 H^2 = \frac{8\pi G}{3}\biggr(\rho - \frac{\rho^2}{\rho_c}\biggl)\;.
\end{eqnarray}
\par
Let us introduce in the Rastall's theory a fluid with an equation of state of the type \cite{gabriel},
\begin{eqnarray}\label{eos}
p \ = \  \alpha\rho + \omega\rho^2\;.
\end{eqnarray}
For our purposes, such equation of state is purely phenomenological. But, it is in principle possible to evoke a Bose-Einstein Condensation (BEC) origin, even if some
details must be worked out in order to have a consistent scenario.
With this equation of state,
the equations of motion are:
\begin{eqnarray}
3\frac{\dot{a}^2}{a^2} &=& 4\pi G \left \{ [ (3 - \gamma) + 3\alpha(\gamma - 1) ] \rho + 3\omega(\gamma - 1)\rho^2 \right \}\;, \\
\dot{\rho}  &=&  -6\frac{\dot{a}}{a}\rho\frac{(1 + \alpha + \omega\rho)}{\left [(3 - \gamma) + 3\alpha(\gamma - 1) + 6\omega(\gamma - 1)\rho\right ]} \;.
\end{eqnarray}
\par
Let us consider $\alpha \neq -1$. The conservation equation takes then the form,
\begin{eqnarray}
 \rho^{A}(1 + \alpha + \omega\rho)^{\frac{B}{\omega}} = \lambda a^{-6}\;,
\end{eqnarray}
with the definition,
\begin{eqnarray}
A \equiv \frac{(3 - \gamma) + 3\alpha(\gamma - 1)}{1 + \alpha}\;, \qquad \frac{B}{\omega} \equiv \frac{(7\gamma - 9) + 3\alpha(\gamma - 1)}{1 + \alpha}\;,
\end{eqnarray}
where $\lambda$ is a \textit{positive} integration constant.
\par
The existence of the pressure term, with the equation of state (\ref{eos}), in the modified Friedmann's equation allows to have
a term similar to that predicted by LQC.
This modified Friedmann's equation obtained from the Rastall's theory leads to many new type of solutions, describing interesting cosmological scenarios. Two of them, of particular interest, are:
\begin{enumerate}
\item Soft singular solutions.
\item Non-singular, bounce solution.
\end{enumerate}
\par
One example of a soft singular solution is given when we choose $\alpha \neq -1, \; \gamma = 3/2$.
The solution reads:
\begin{eqnarray}
 \rho = -\frac{1 + \alpha}{2\omega} \left(1 \pm \sqrt{1 + \frac{4\omega\lambda}{(1 + \alpha)^2} a^{-4}} \right)\;,
a \propto t^{\frac{1}{2}}, \qquad H \propto t^{-1}\;,
\end{eqnarray}
Remark that this solution gives a behaviour for the scale factor identical to the radiative phase of the
standard model.
However, $a$ can not go to zero if $\omega$ is negative, otherwise the density becomes imaginary.
The cosmic evolution begins necessarily at a $t = t_0 \neq 0$.
This curious solution may contain a geodesic singularity.
In some sense it remembers the sudden singularity solution \cite{barrow} but for the primordial universe.
It may also have some connections with the singularities found in the DBI context in reference \cite{kame}: singularities that
can be crossed by a pointlike object.
This may lead also to some possible similarities with the cosmic genesis project developed in reference \cite{lukash}.
\par
Let us consider now $\alpha \neq -1, \; \gamma = \frac{3(\alpha + 3)}{3\alpha + 7}$.
The solution reads:
\begin{eqnarray}
a \ = \ \left [ \frac{2\pi G \lambda(1 + \alpha)}{\gamma - 1}t^2 - \frac{\omega\lambda}{2(1 + \alpha)} \right ] ^{\gamma - 1}\;.
\end{eqnarray}
If $\alpha > -1$, then  $\gamma - 1$ and $1 + \alpha$ is positive. If $\omega < 0$, the scale factor varies from infinity ($t \to -\infty$) to infinity ($t \to +\infty$) with a minimum value  $a = \left[\frac{|\omega|\lambda}{2(1 + \alpha)}\right ]^{\gamma - 1}$ and the energy density varies from zero to zero with the maximum value $\rho = 2(1+\alpha)/|\omega|$.
Hence, there is no singularity in the cosmic evolution for this case. One important advantage of such approach to the LQC effective Friedmann is that we have a full set of covariant equations, what may allow to inspect the model at perturbation level. Of course, we can not assure that all the structure of the underlying quantum model is present in this classical
approach; but, some hints on the behaviour of the corresponding quantum model at linear perturbation level may be obtained.
\par
Remark that, in order to have a singularity free solution, it is necessary to have a connection of the equation of state parameter
$\alpha$ and
the Rastall's parameter $\gamma$.
But, since $\gamma$ quantifies the violation of the usual conservation law, it is not impossible that it could depend on
the kind of fluid considered.

\section{Application II: The present universe}

The present universe is very well described by the $\Lambda$CDM model.
However, this description has some important drawbacks.
Among then, there is the excess of power in the matter spectrum at
small scales.
Can the Rastall's cosmology give some new insight on these problems
preserving the success of the $\Lambda$CDM model?
\par
Let us consider a two fluid model, when representing dark matter (and also baryons) and the other the dark energy component under the form of a cosmological term \cite{grupo1}:
\begin{eqnarray}
R_{\mu\nu} - \frac{1}{2}g_{\mu\nu}R &=& 8\pi G\biggr\{T^m_{\mu\nu} + T^x_{\mu\nu} - \frac{\gamma - 1}{2}g_{\mu\nu}(T^m + T^x)\biggl\},\\
{T^{\mu\nu}_x}_{;\mu} &=& \frac{\gamma - 1}{2}(T_m + T_x)^{;\nu},\\
{T^{\mu\nu}_m}_{;\mu} &=& 0.
\end{eqnarray}
The structure displayed above takes into account the fact that it is necessary to have the usual conservation law for one of the fluids (the pressureless one representing at this stage baryons and dark matter) in order to allow structure formation through gravitational collapse.
The second fluid, represented by the subscript $x$, obeys the vacuum equation of state, $p_x = - \rho_x$.
\par
The equations of motion are
\begin{eqnarray}
H^2 &=& \frac{8\pi G}{3}\biggr\{(3 - 2\gamma)\rho_x + \frac{-\gamma + 3}{2}\rho_m\biggl\},\\
\dot\rho_m + 3H\rho_m &=& 0,\\
(3 - 2\gamma)\dot\rho_x &=& \frac{\gamma - 1}{2}\dot\rho_m,\\
\rho_m &=& \frac{\rho_{m0}}{a^3},\\
\rho_x &=& \frac{\rho_{x0}}{3 - 2\gamma} + \frac{\gamma - 1}{2(3 - 2\gamma)}\rho_m.
\end{eqnarray}
\par
Combining these expressions we obtain:
\begin{eqnarray}
H^2 = \frac{8\pi G}{3}(\rho_{x0} + \rho_m),\\
2\frac{\ddot a}{a} + \biggr(\frac{\dot a}{a}\biggl)^2 = 8\pi G \rho_{x0}.
\end{eqnarray}
It is the same background dynamics of the $\Lambda$CDM model. Hence, for this model, the so-called kinematical observational tests, based on the dynamics of the
background, are satisfied in exactly the same way as in the $\Lambda$CDM model.
\par
At linear perturbation level, using the standard calculations in the synchronous coordinate condition, we find the following equation for the evolution of the density contrast of the pressureless component \cite{grupo1},
\begin{eqnarray}
\ddot\delta_m + 2\frac{\dot a}{a}\dot\delta_m - 4\pi G\rho_m\delta_m = 0.
\end{eqnarray}
Again, it is the same perturbed equation as in the standard $\Lambda$CDM model. Hence, the perturbation tests based on the linear approximation, like the matter power spectrum, are identical
to the $\Lambda$CDM model.
\par
But now, there is the new relation relating the fluctuations of dark matter to the fluctuations of dark energy,
\begin{eqnarray}
\delta\rho_x = \frac{\gamma - 1}{2(3 - 2\gamma)}\delta\rho_m.
\end{eqnarray}
Dark energy agglomerates, even if it is in the form of a vacuum energy term.
This may modify the $\Lambda$CDM scenario at non-linear level, and may lead to some new scenarios at this regime where the standard model faces some difficulties.
\par
In the previous model, it has been introduced a cosmological constant
and a dark matter component.
It is possible to have realistic variations around this scenario as in the usual standard model?
\par
Let us consider the same structure as before but with
a more general equation of state for the dark energy component \cite{grupo2}:
\begin{eqnarray}
p_x = w_x\rho_x.
\end{eqnarray}
The previous case implies $w_x = - 1$.
\par
The equations of motion are now,
\begin{eqnarray}
\left(\frac{\dot{a}}{a}\right)^{2} &=& \frac{4\pi G}{3}\left\{\rho_{x}\left[3-\gamma - 3(1-\gamma)w_{x}\right] + (3-\gamma)\rho_{m}\right\}\;,
\\
\frac{\ddot{a}}{a} &=& \frac{4\pi G}{3}\left\{\left[3(\gamma-2)w_{x} -\gamma\right]\rho_{x}  - \gamma\rho_{m}\right\}\;,\\
\dot{\rho}_{x} + 3\frac{\dot{a}}{a}(1+ w_{x})\rho_{x} &=& \frac{\gamma-1}{2}\left[(1-3w_{x})\dot{\rho}_{x}  + \dot{\rho}_{m}\right]\;,
\\
\label{rhomeq}\dot{\rho}_{m} + 3\frac{\dot{a}}{a}\rho_{m} &=& 0\;.
\end{eqnarray}
The density parameter are now given by:
\begin{eqnarray}
\rho_x &=& \rho_{de0} a^{-\frac{6(1+w_x)}{2-(\gamma-1)(1-3w_{x})}} + \frac{(1-\gamma)\rho_{m0}}{2w_x + (\gamma-1)(1-3w_x)}a^{-3}\;,
\\
\rho_{m} &=& \frac{\rho_{m0}}{a^{3}}\;.
\end{eqnarray}

\begin{center}
\begin{figure}[!t]
\begin{minipage}[t]{0.3\linewidth}
\includegraphics[width=\linewidth]{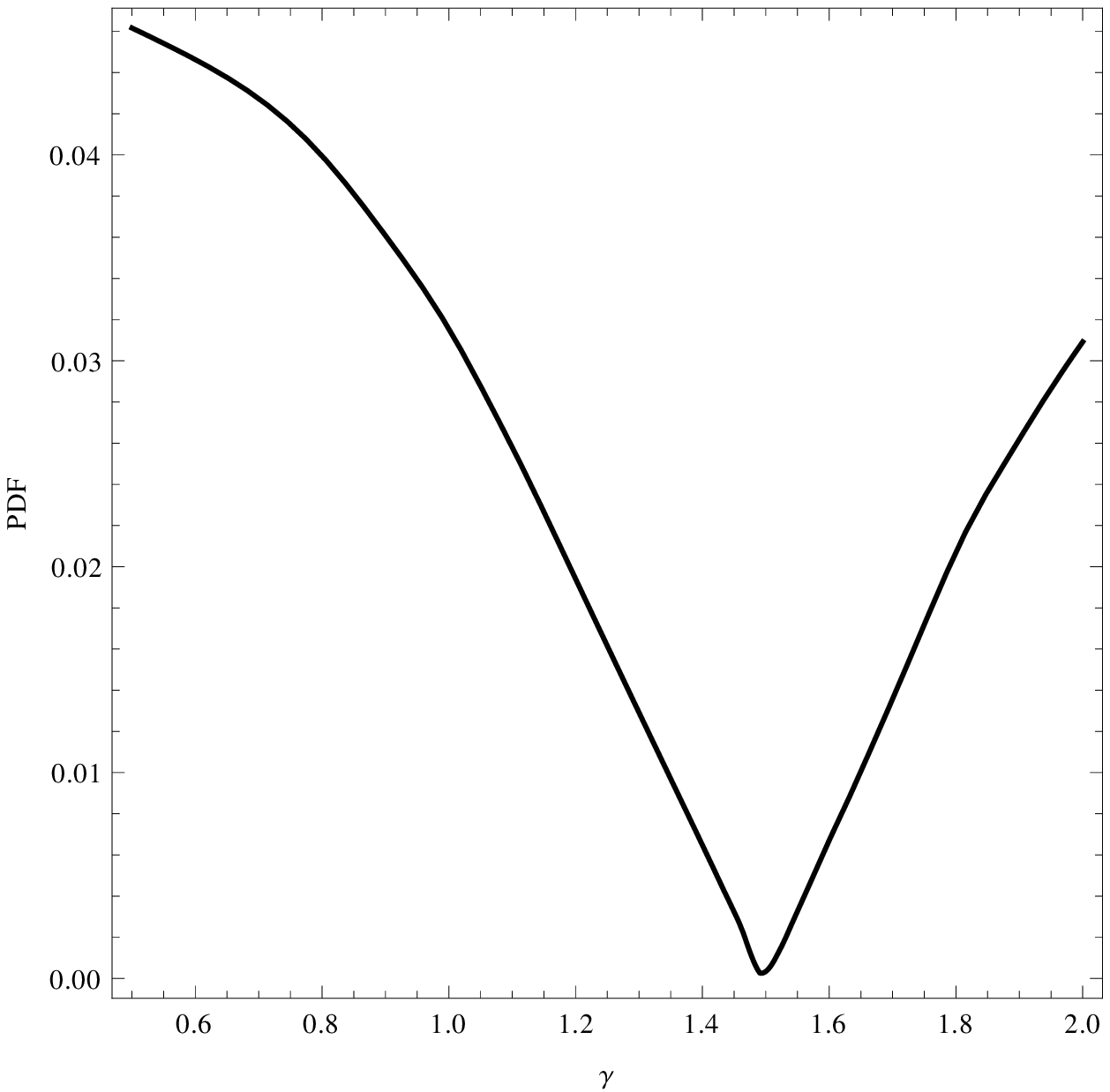}
\end{minipage} \hfill
\begin{minipage}[t]{0.3\linewidth}
\includegraphics[width=\linewidth]{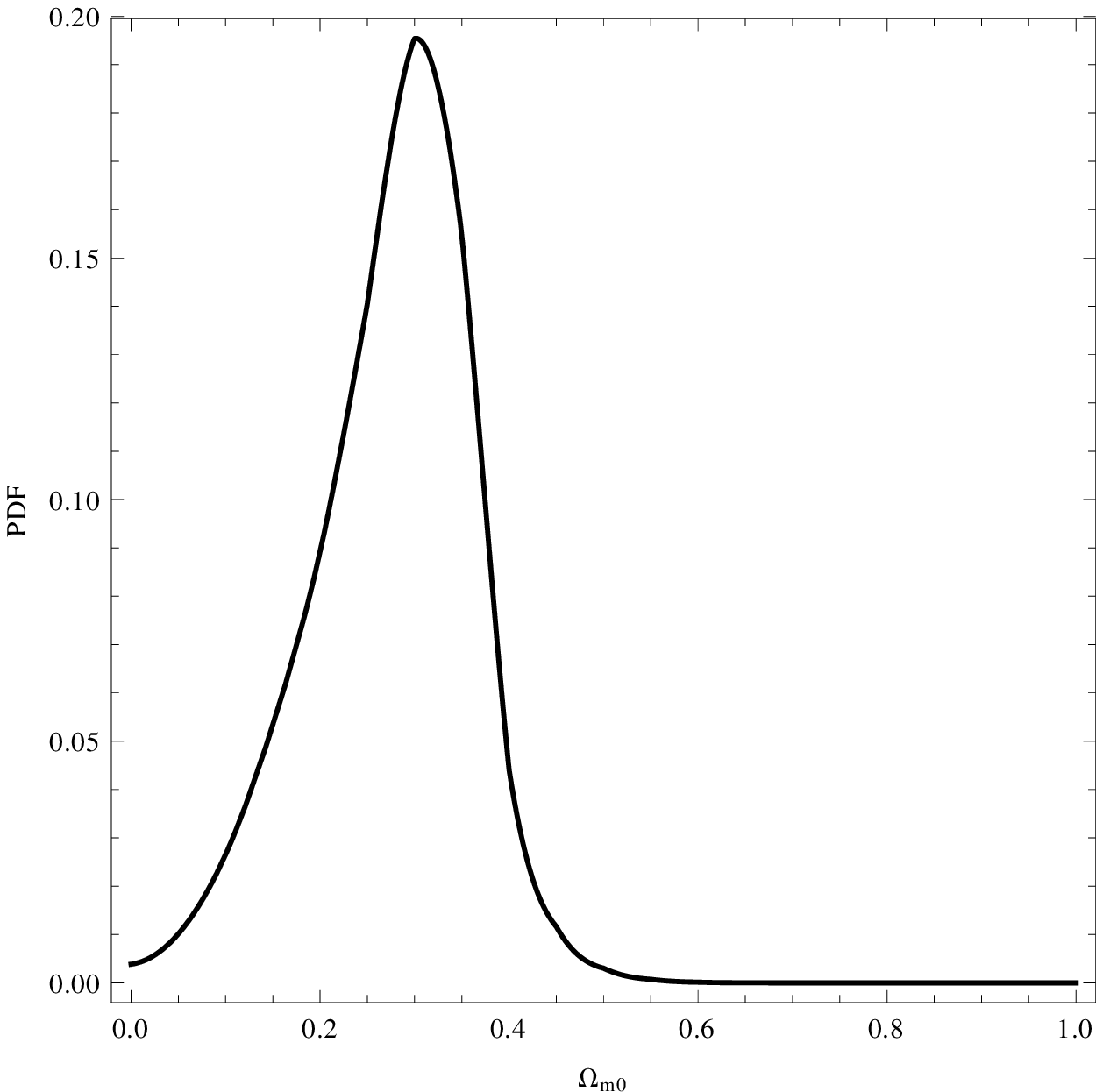}
\end{minipage} \hfill
\begin{minipage}[t]{0.3\linewidth}
\includegraphics[width=\linewidth]{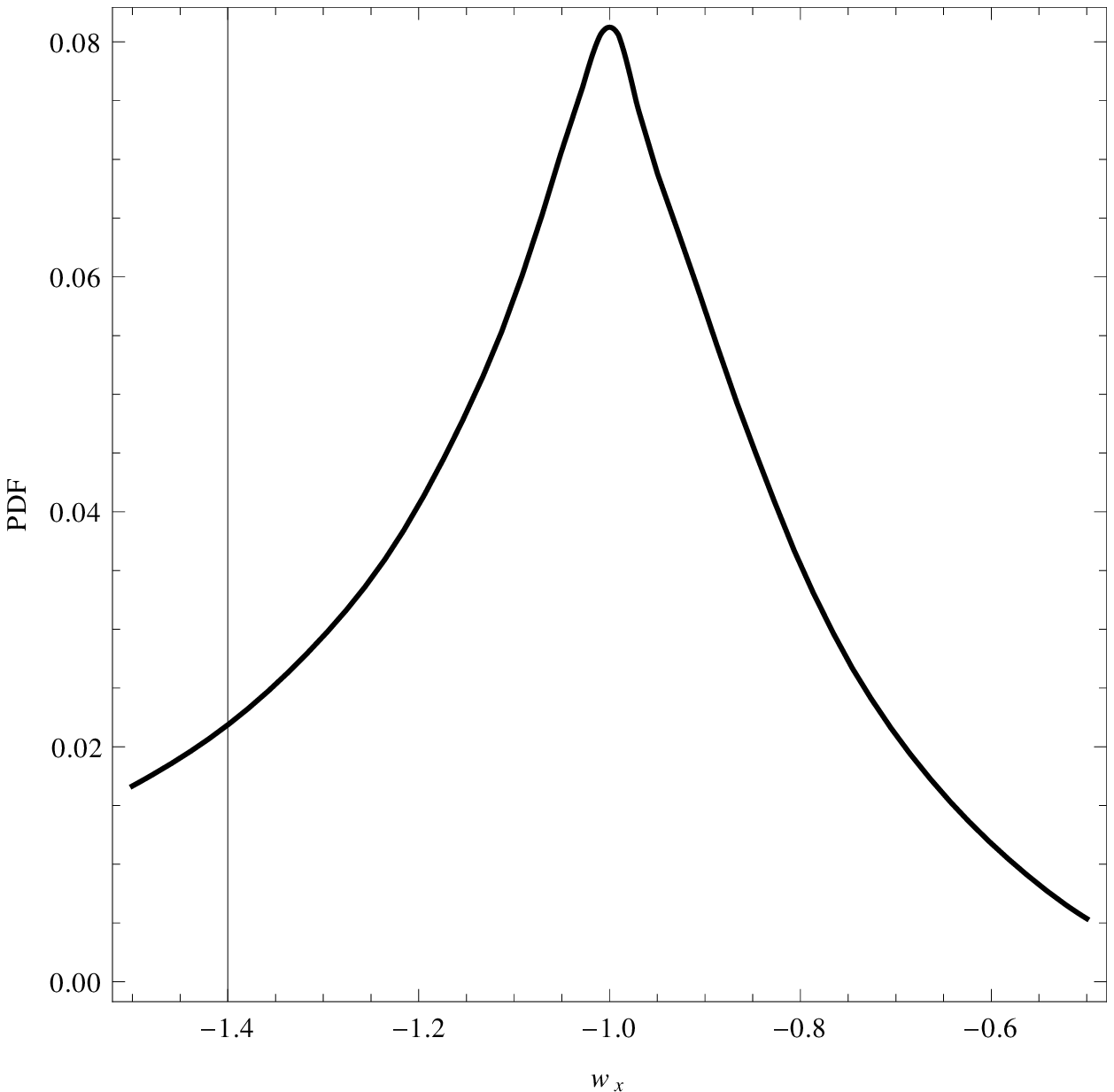}
\end{minipage} \hfill
\caption{Probability distribution function for the parameters of the Rastall's cosmology
(at the left panel $\gamma$, in the middle panel the matter density parameter and at the right panel
the dark energy equation of state parameter) obtained by using supernova type Ia data.}
\end{figure}
\end{center}

\begin{center}
\begin{figure}[!t]
\begin{minipage}[t]{0.3\linewidth}
\includegraphics[width=\linewidth]{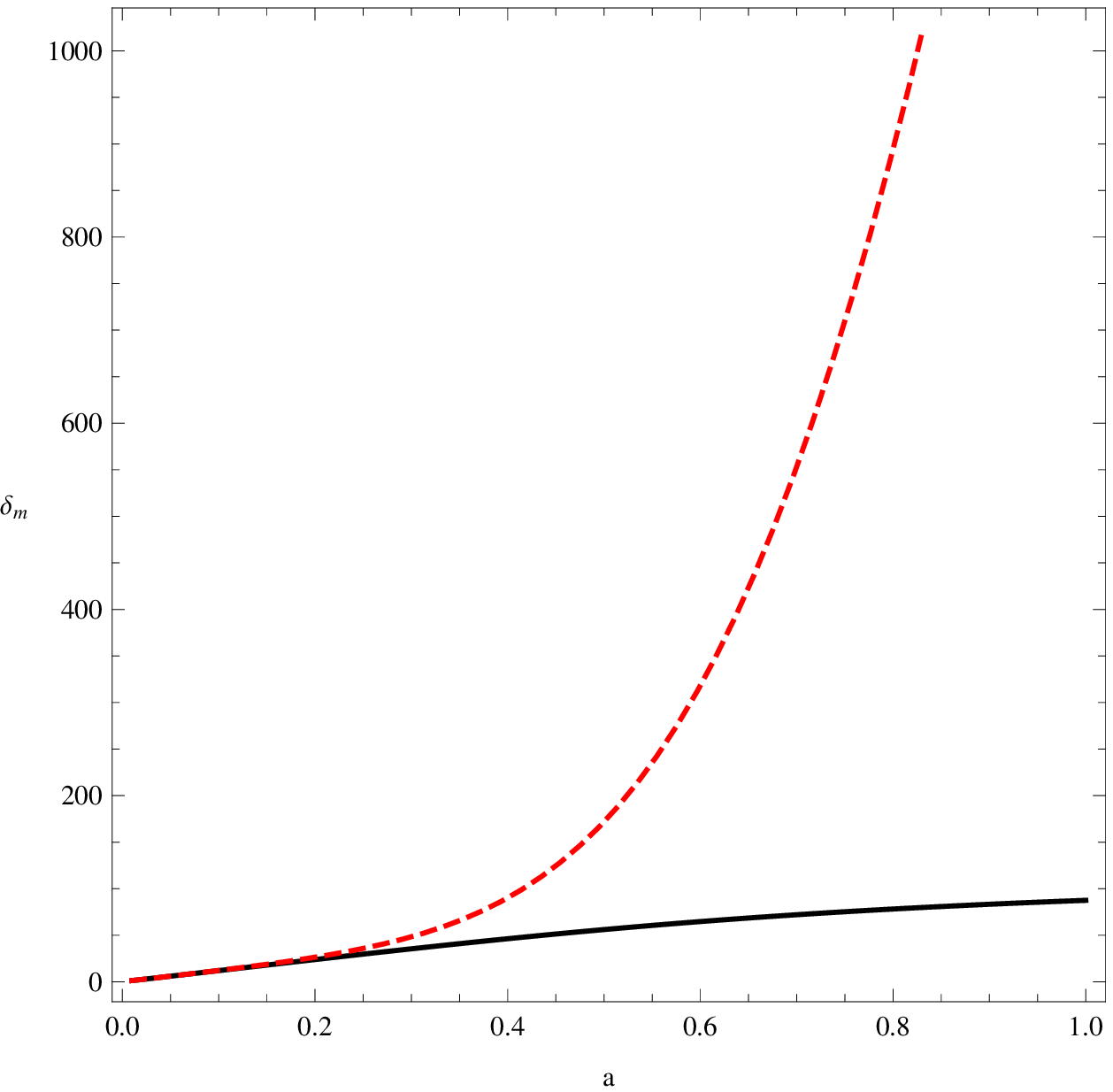}
\end{minipage} \hfill
\begin{minipage}[t]{0.3\linewidth}
\includegraphics[width=\linewidth]{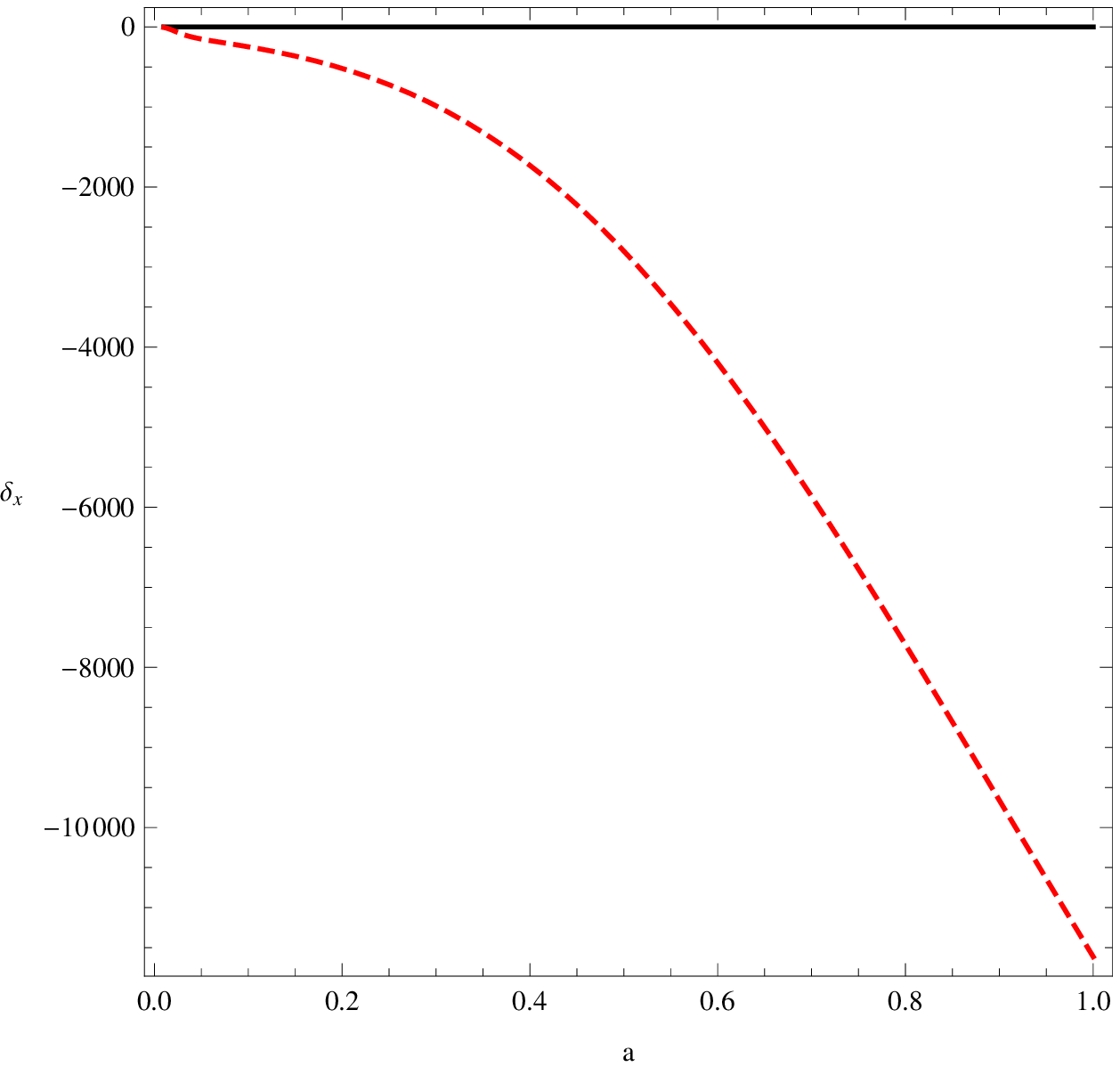}
\end{minipage} \hfill
\caption{Typical behaviour of the time evolution of the matter (left panel) and dark energy (right panel) density contrasts in the Rastall's cosmology with $\omega_x \neq - 1$ (red curve) and with
$\omega_x = - 1$ (black curve).}
\end{figure}
\end{center}

The situation now changes drastically. In principle, the background tests can be satisfactorily satisfied for some range of values
of $\gamma$ and $\omega_x$, as shown in figure 1. But, at perturbation level the constraints are much more strict:
a small deviation even of the order of $10^{-4}$ from $w_x = - 1$ is catastrophic at perturbation level.
Essentially, only the case $w_x = - 1$ survives. Hence, in the Rastall's cosmology the comparison to observation seems to
single out the $\Lambda$CDM model, mainly due to the perturbation behaviour even at linear level.

\section{Conclusions}

Rastall's theory is a possible mechanism to introduce in a gravity theory quantum effects in a classical, effective approach. Rastall's theory touches
one of the cornerstone of the General Relativity theory: the conservation of the energy-momentum tensor. Now, the divergence of
the energy-momentum tensor is proportional to the gradient of the Ricci curvature scalar.
\par
One natural application of the Rastall's theory, viewed as an effective theory implementing classically the general features of
quantum effects, is the primordial universe. With a suitable equation of state, it is possible to reproduce in the context of
the Rastall's cosmology, the general behaviour predicted by Loop quantum cosmology. Singularity-free cosmological models are
obtained. Some types of soft singularities are also displayed.
\par
In applying the Rastall's cosmology to the present universe, and using a model containing pressureless matter and a cosmological term, it is possible to keep the success of the $\Lambda$CDM model, but introducing new features
at non-linear perturbation level.
This {\it modified $\Lambda$CDM model} is, in some sense, unique, since any deviation of this configuration, by generalizing the equation of the state of the dark energy component, leads to a complete disagreement with observations, mainly with respect to the power spectrum data.
\newline
\vspace{0.5cm}
\newline
{\bf Acknowledgements:} We thank FAPES (Brazil) and CNPq (Brazil) for partial financial support. J.C.F acknowledges also support by
“FONDECYT-Concurso incentivo a la Cooperaci´on Internacional”
No. 1130628.

%%%%%%%%%%%%%%%%%%%%%%%%%%%%%%%%%%%%%%%%%%%%%%%%%%%%%%%%%%%%%%%%%%%%%%%%%%%%%%%%%%%%%%%%%%%%%%%%%%%%%%

\end{document}